# Masses of Astrometrically-Discovered and Imaged Binaries: G 78-28AB and GJ 231.1BC

Steven H. Pravdo
Jet Propulsion Laboratory, California Institute of Technology
306-431, 4800 Oak Grove Drive, Pasadena, CA 91109; spravdo@jpl.nasa.gov

Stuart B. Shaklan
Jet Propulsion Laboratory, California Institute of Technology
301-486, 4800 Oak Grove Drive, Pasadena, CA 91109, shaklan@huey.jpl.nasa.gov

Sloane J. Wiktorowicz,
Division of Geological & Planetary Science, California Institute of Technology
Pasadena, CA 91101, sloane@gps.caltech.edu

Shri Kulkarni
Division of Physics, Mathematics, and Astronomy, California Institute of Technology
Pasadena, CA 91101, srk@astro.caltech.edu

James P. Lloyd
Department of Astronomy, Cornell University, Ithaca NY 14853, jpl@astro.cornell.edu

Frantz Martinache
Department of Astronomy, Cornell University, Ithaca NY 14853,
frantz@astro.cornell.edu

Peter G. Tuthill,
School of Physics, University of Sydney, NSW 2006 Sydney, Australia,
gekko@physics.usyd.edu.au

&
Michael J. Ireland
Division of Geological and Planetary Sciences, California Institute of Technology,
Pasadena, CA 91125, mireland@gps.caltech.edu

**ABSTRACT**

The Stellar Planet Survey (STEPS) is an ongoing astrometric search for giant planets and brown dwarfs around a sample of ~30 M-dwarfs. We have discovered several low-mass companions by measuring the motion of our target stars relative to their reference frames. The highest mass discovery thus far is G 78-28B, a companion to the M-dwarf G 78-28A. The orbital period is $4.18 \pm 0.03$ y, the system mass is $0.565 \pm 0.055$ $M_\odot$, and the semi-major axis is $2.19 \pm 0.10$ AU. Imaging observations with the Keck laser guide star adaptive optics (LGSAO) and the Palomar AO instruments resolved the system and also yielded *JHK*-band delta magnitudes. We use the orbital solution, light ratios, and mass-luminosity relationships to derive component masses of $M_A = 0.370 \pm 0.034$ $M_\odot$ and $M_B = 0.195 \pm 0.021$ $M_\odot$. G 78-28B is of type M4 V based upon its colors and mass. We also discovered GJ 231.1C, a companion to GJ 231.1B, with STEPS and imaged the companion with LGSAO and Palomar AO, but the orbital period is longer than our observing baseline; thus the system parameters are less constrained. In GJ 231.1BC the masses are $M_B = 0.25 \pm 0.06$ $M_\odot$ and $M_C = 0.12 \pm 0.02$ $M_\odot$. The inferred spectral type of GJ 231.1C is M5 V. We demonstrate the results of the current state of mass estimation techniques with our data.

# 1. INTRODUCTION

The Stellar Planet Survey (STEPS) is an astrometric search for low mass companions to M-dwarfs. Astrometry provides a direct measurement of stellar mass because the full three-dimensional orbit is observed. Measurements of mass test and assist the development of the models based upon parameters such as age and metallicity. Determining an accurate mass thus deepens our understanding of the fundamental physics of stars and substellar objects. Another direct benefit is to advance our knowledge of the mass-luminosity relationships (MLRs). At present there are no extant observational MLRs for brown dwarfs (BDs) and the MLR for stars at the bottom of the main sequence is based upon only 10 objects (Henry et al 1999). We have already made several mass measurements of companions to M-dwarfs with the Stellar Planet Survey (STEPS, Pravdo, Shaklan, Henry, & Benedict 2004, Pravdo, Shaklan, Lloyd, & Benedict 2005, and Pravdo, Shaklan, & Lloyd 2005). In each case the combination of astrometry and imaging resulted in conclusions about the masses of the components that could not have been reached by either technique alone.

Herein we report on two more of the ~30 targets that at the beginning of the STEPS program were considered "single" stars. We astrometrically discovered companions around both G 78-28 and GJ 231.1B and we later confirmed the existence of the companions with laser guide star adaptive optics (LGSAO) and Palomar adaptive optics (AO) imaging observations. We present these results and discuss how they contribute to MLRs in particular and stellar knowledge in general.

# 2. OBSERVATIONS AND RESULTS

## 2.1 Astrometry

### *2.1.1 G 78-28AB*

G 78-28 (=G 95-22, LTT 17492, GJ 3213) is an M3 dwarf with the properties listed in Table 1. The parallax has not been determined trigonometrically and is estimated from the spectral type and colors as 66 ± 13 *mas* (Gliese & Jahreiss 1991) and 59 *mas* (with large error; Reid, Hawley, & Gizis 1999). We observed it from Dec. 1997 through Jan. 2005 with the STEPS instrument (astrometric bandpass 550-750 nm) mounted at the Cassegrain focus of the Palomar 200" (5-m) telescope. The first observation was Dec. 21.3., 1997 = JD 2450803.8. Pravdo et al (2004, 2005) give more detailed descriptions of the instrument and data analysis.

Table 2 shows the results of our measurements of parallax and proper motion. Our parallax is measured relative to the in-frame reference and should be corrected for the reference frame's finite distance by the addition of ~2 ± 1 *mas* for average fields at this galactic latitude and apparent magnitude (van Altena, Lee, & Hoffleit 1995). The result is $\pi = 54.4 \pm 1.0$ *mas*. Our value is on the lower boundaries of the prior measurements, consistent with the fact that the primary star is now 0.2-0.3 magnitudes fainter in *VJHK* than when it was believed to be the origin of all the light. Our proper motion values are consistent with prior results where the error bars on the prior results are estimated from the variation among past observers (Luyten 1979, CMC 1999, Salim & Gould 2003). Proper motions of reference stars in frames can suggest false accelerations of the target stars, limiting the accuracy of orbits with periods long compared to the observational baseline. However, the effect is negligible for STEPS, as the corrections are only hundredths of *mas* for baselines of ~10 years.

G 78-28 has a periodic astrometric signal after subtraction of parallax and proper motion from the total motion, indicating the presence of the companion, G 78-28B. Figs. 1 and 2 shows the astrometric data superimposed on an orbit with an acceptable fit. Our error estimates comprise the uncertainty due to the Poisson errors derived from the standard error of the mean of a set of 10-20 exposures each night added in quadrature to 1.5 *mas* systematic errors. The latter could be due to unmodeled real motions (e.g., other companions) or currently unmeasured instrumental errors. We use the Monte Carlo technique to determine the 1-sigma confidence limits in our observed parameters via the method described in Lampton, Margon, & Bowyer (1976) for multi-parameter estimation (see Pravdo et al. 2004 for further details). Table 2 lists the orbital parameters.

*2.1.2 GJ 231.1 BC*

We observed GJ 231.1B (=G 106-36, HD 43587B) over the same approximate span as G 78-28, but starting one day earlier, JD 2450802.8. Table 1 also lists its previously known properties. The parallax value for the companion GJ 231.1A (an F9V star) is 50 ± 9.6 *mas* (van Altena, Lee, & Hoffleit 1995), consistent with the Reid, Hawley, & Gizis (1995) combined trigonometric and photometric parallax for GJ 231.1B of 50 ± 10 *mas*. The proper motion and position angle (PA) values in Table 1 are the mean values of the prior measurements listed, and the errors are the standard deviations. The STEPS absolute parallax is 55.2 ± 1.0 (Table 2). This and the measured proper motion are consistent with the prior results.

The astrometric fits to the GJ 231.1B motion indicate the presence of the companion GJ 231.1C. The residual or systematic error in the fits is 1.7 *mas* and arises in part from the variation with orbital phase in the shape of the point-spread function (PSF) of GJ 231.1BC. We now know (see following section) this PSF contains a bright companion with a separation that is ~⅓ the size of the PSF of an unresolved source in good seeing. These variations add noise to the fitting process.

Fig. 3 shows the allowed orbital periods for GJ 231.1BC versus system eccentricity. The minimum period is ~25 y. The period is <200 y for eccentricity $e < 0.7$. Fig. 4 shows the data superposed upon a possible orbit with ~100 y period. The relatively short temporal baseline compared with the minimum period does not allow us to usefully restrict the dynamical mass. Fig. 3 shows values in models for which the total mass is limited by the MLRs (§3.3).

**2.2 Laser Guide Star Adaptive Optics Imaging**

*2.2.1 Observations*

Observations of G 78-28 and GJ 231.1B were taken on Jan. 5, 2005 with the laser guide star adaptive optics (LGSAO) system at the Keck II telescope (Wizinowich, *et al*. 2004) and the narrow camera mode of the Near-Infrared Camera 2 (NIRC2). The laser excites a region of sodium atoms at an altitude of roughly 90 km, and the resulting emission is bright enough to allow for correction of the lower atmosphere by the rest of the adaptive optics train. The advent of LGSAO allows fainter objects to be observed with adaptive optics, because a bright star (V magnitude ≈ 13) is no longer necessary for high-order correction. However, since the laser and the observed sodium emission sample the same volume of air, low-order, tip-tilt correction requires another source. This is satisfied by a nearby, fainter star (V magnitude ≈ 16), because low-order correction requires fewer photons.

G 78-28 was observed in *J*, *H*, and *Kp* bands, and only the *H* band data were used for astrometry (710 co-added images of 0.2 sec each, see Fig. 5). The *Kp* band is centered at 2.124 μm with a bandwidth of 0.351 μm[1]. The binary separation was too small for the poor *J* band correction to be useful. While the *Kp* band correction was excellent, the images were unfortunately taken in a mode that allowed the sky to rotate on the detector. This made accurate PA determination difficult. We obtained photometry in all three bands.

GJ 231.1B was observed in *J* (130 co-added images of 0.5 sec each) and *Kp* bands (550 co-added images of 0.5 sec each, see Fig. 6); astrometry and photometry were determined for both. Even though the *J* band data are less extensive, astrometry between *J* and *Kp* bands agrees to 0.5 *mas* separation and 0.20 degrees PA. We summarize the data from both sources in Table 3. Fig. 2 shows the LGSAO data with the other data and the orbital model.

*2.2.2 Data Reduction*

G 78-28A and B are partially resolved; B sits on a substantial gradient of the A PSF (Fig 5). We estimated the value of the A PSF at the location of the B peak by measuring the value of the A PSF at several radial distances equal to the nominal separation of AB. GJ 231.1B and C are well resolved (Fig. 6) but C sits on a significant background due to the adaptive optics halo around B.

The background was removed by taking advantage of the circular symmetry of the AO halo. We rotated the image of GJ 231.1BC about the peak of B and subtracted it from the non-rotated image. This resulted in positive and negative images of C with a nominally flat background except for evidence of the AO side lobes close to the B image. The halo appears to be removed to ~1% of the B peak.

We measured the separations and PAs using the *Kp* images because the G 78-28 and GJ 231.1 PSFs are similar. The PSFs in J of the GJ231.1 image are ~2 times broader than for G 78-28, making image subtraction problematic. The PSF fitting region was 80 *mas*. We fitted the GJ 231.1B PSF to the core of the G 78-28A PSF while masking the pixels at and around the G 78-28B image. Fitting parameters were position and amplitude. We then subtracted the shifted and scaled GJ 231.1B image from the G 78-28A image, resulting in a nominally flat background clearly showing a positive peak at G 78-28B and a negative peak at GJ 231.1C. Finally, we fitted a spline to the residuals and measured the peak location. The measurements of the pixel scales are described in the appendix.

---

[1] See http://alamoana.keck.hawaii.edu/inst/nirc2/Manual/ObserversManual.html

## 2.3 Palomar Adaptive Optics Imaging

We observed G 78-28 in the *H* band with the Palomar 200" Adaptive Optics (AO) System (Troy et al. 2000) as part of a program to explore precision calibration of AO images with the application of non-redundant aperture-masking interferometry. The difficulties described in Section 2.2.2 regarding the removal of the PSF and extraction of the astrometric and photometric measurements are inherent to the images produced by AO systems. The fluctuation of the unstable AO PSF limits the precision and sensitivity of AO observations. Substantial progress has been made in solving this problem by exploiting simultaneous differential measurements in polarization (Apai et al. 2004; Potter 2003; Perrin et al. 2004) or wavelength (Close et al. 2005; Marois et al. 2005). These techniques are useful in cases where there is a large differential signal, but do not address the generic problem of the fidelity of AO imaging.

An alternative approach to the exploitation of the coherent wavefront provided by an AO system is the application of aperture-masking interferometry (Tuthill et al. 2000) instead of conventional imaging. For these observations, we placed a 9-hole mask in the Lyot stop of the PHARO camera (Hayward et al. 2001) with 50-cm projected hole diameters and a longest baseline of 415 cm. The hole positions are chosen to maximize Fourier coverage and transmission, while maintaining non-redundancy to preserve closure-phase (Haniff et al. 1987, Readhead et al. 1988). The resulting interferogram is recorded in the image plane. This interferogram records all 36 pair-wise fringes from the 9 holes in the mask.

The advantages of this approach are several-fold. The AO PSF instability is a result of the fluctuations in the residual atmospheric phase and AO system calibration errors. The preservation of closure phase by the non-redundancy allows the use of self-calibration techniques (Cornwell 1989), thus rejecting these residual phase errors. The calibration problem is also simplified to a well-posed problem of calibrating the visibility of a single interferometer baseline at one time, rather than the ill-posed inverse problem of deconvolution of an image with an unknown PSF. Finally, an interferometric approach enables "super-resolution" if the calibration is sufficiently accurate. In comparison with uncompensated aperture masking interferometry (Tuthill et al. 2000; Nakajima et al. 1989), AO provides stabilization of the fringes enabling long integration times and therefore reach to fainter targets.

The data are calibrated with observations of a nearby star. Care must be taken to select a source that is of similar brightness to both the AO wavefront sensor operating in the red optical, and the science camera operating in the infrared to ensure a comparable wavefront correction and signal-to-noise ratio. The data is dark-subtracted, flat-fielded, and analyzed with a custom software pipeline written in IDL. The pipeline outputs a bispectrum in OIFITS format (Pauls et al. 2005). A binary model is fit to the bispectrum. Although the binary signal is apparent in visibility amplitude through a power spectrum, in practice we have found that the visibility amplitude calibration is poor, and superior results are achieved with a fit to the closure phase alone. Presumably this is because the visibility amplitude calibration is susceptible to the same fluctuations in seeing and AO performance between source and calibrator that plague conventional imaging with AO. For the observations reported here we neglect the visibility amplitude and the model is fit to the phase of the bispectrum.

G 78-28 was observed in four observing runs in December 2003, September 2005, December 2005 and February 2006. No fringes were detected in the September 2005 observations due to poor seeing. At *V* magnitude 12.4, G 78-28 is near the performance limit of the natural guide star AO system, requiring the AO system to operate at reduced bandwidth. The fringe stability is therefore a sensitive function of the atmospheric turbulence. Median closure phase scatter were 3.5, 0.6 and 1.2 degrees in the Dec. 2003, Sep. 2005, and Feb. 2006 data sets, respectively. The Sep. 2005 (JD 2453632.9) data are in the best seeing, and fortuitously, at the largest separation. The bispectrum model fit typically had reduced $X^2 > 1$, attributed to a systematic effect. We added a systematic error to achieve $X^2 = 1$, and determine the confidence intervals. Starting in Feb. 2006, we increased the number of times that we cycled between source and calibrator, and this eliminated the need for the added error. The resulting final closure phase errors were 4.3, 2.0 and 1.2 degrees on the three datasets.

The resulting extraction of astrometric parameters is a sensitive function of the orbital separation. In cases where the binary is well resolved, the solution is unambiguous. On JD 2453779.7 the separation of 96 *mas* (1.45 λ/D) is well resolved and the resulting likelihood function is unambiguously fit with a unique separation and contrast (Fig. 7a). However, on JD 2453632.9 the separation is only 41 mas (0.62 λ/D). Even though this is well below the conventional resolution limit of the telescope, the binary is well detected, but there is degeneracy between the separation and contrast ratio. The likelihood contours shown in Fig. 7 approximately define a locus of constant closure phase. For small separations and brightness ratios, closure phase is proportional to brightness ratio and separation cubed. Since the contrast ratio is well-constrained by the better-resolved observation, we adopt the JD 2453779.7 contrast ratio for the other observations. The resulting parameters are shown in Table 4. The *H*-band magnitude difference, 1.285 ± 0.023, is consistent with LGSAO result at better than 1 σ. Fig. 2 shows the Palomar AO orbital data with the other data and the orbital model.

GJ 231.1B was observed on JD 2453779.7. It was well resolved and detected with conventional AO imaging. At *V* = 13.3 magnitude it is near the performance limit of the natural guide star AO system. Due to the fluctuating wavefront quality, we found it necessary to take about 100 short exposure images in *H* and $K_s$ bands, and only use the images with highest Strehl ratio to extract the brightness ratios. The separation, PA, and delta magnitudes for *H* and $K_s$ are shown in Table 4. The *K*-band delta magnitudes from LGS and Palomar AO are consistent with each other. The positional data are plotted in Figure 4.

### 3. DISCUSSION

#### 3.1 G 78-28 AB

The G 78-28AB system is 18.4 ± 0.3 pc from the Sun. The composite light was spectrally classified as a dwarf M3 (e.g. Reid, Hawley, & Gizis 1995). It is a moderately active X-ray star (Hünsch et al. 1999) with about ⅓ the X-ray luminosity of GJ 802Ab, a newly discovered system with a close companion (Pravdo, Shaklan, & Lloyd 2005). Although there are only small numbers of identified close binaries such as these, a correlation between M-dwarf with close companions and X-ray emission may emerge as the binarity of heretofore "single" M dwarfs is discovered. Its (*U,V,W*) space velocity measured by Reid, Hawley, & Gizis (19915) is consistent with it being a member of their local volume-complete sample of M dwarfs.

We measure the dynamic total mass in the astrometric fits to be $0.60 \pm 0.09$ $M_\odot$. We also find that $\alpha/a = f - \beta = 0.195 \pm 0.025$, where $\alpha$ is the photocentric axis, $a$ is the semi-major axis, $f$ is the secondary mass fraction and $\beta$ is the secondary light fraction. From the $M_V$ of the system and the visible MLRs we calculate $f$ and $f - \beta$ with $\beta$ as the independent variable. The measured $f - \beta$ corresponds to a total mass $<0.51$ $M_\odot$ (eqn. 5b of Henry et al. 1993, eqn. 7 of Henry et al. 1999) or $<0.57$ $M_\odot$ (Delfosse et al. 2000) where the uncertainty in the upper limits is 0.11 $M_\odot$ (from the dispersion given in Henry et al.) and the lower limits are not determinable because the secondary mass becomes too small to be in the applicable range of the MLRs.

Table 5 shows the magnitudes derived for both components in four bands. We derive the delta $V$ between G 78-28A and B of $1.85 \pm 0.28$ mag. from the measured $M_{JHK}$ (Tables 1, 3, 4) and the color-magnitude relationships for M-dwarfs (equations 1a-c of Henry et al.1993). The delta $V$ magnitude corresponds to $0.12 \leq \beta \leq 0.19$ but the measured $f - \beta$ can further constrain it depending upon the MLR. For Henry et al., $0.12 \leq \beta \leq 0.17$. The magnitude range alone constrains the total mass to be $0.508 \pm 0.002$ $M_\odot$ and $f = 0.33 \pm 0.01$. For Delfosse et al, $\beta$ is not further constrained, and the total mass is $0.554 \pm 0.009$ $M_\odot$ and $f = 0.37 \pm 0.02$. The predictions of the MLRs are consistent, once we consider the mass dispersions (Henry et al.), viz., 0.11 $M_\odot$ for the sum. A further detailed discussion of the system masses is in §3.3 where we estimate a consistent value and error based upon the MLRs in several bands.

With $V$-$K$ values of $4.54 \pm 0.05$ and $5.25 \pm 0.28$, the inferred spectral types are M 2.5 and M 4, for A and B, respectively, within a couple tenths of a subtype (e.g., Leggett 1992).

### 3.2 GJ 231.1BC

GJ 231.1A and B are a parallax and proper motion pair (Poveda et al. 1994) consisting of a F9V primary and a M3.5 V secondary (Reid, Hawley, & Gizis 1995) at a distance of $18.1 \pm 0.3$ pc. The secondary is now shown to consist of two stars, B and C. The large separation of B and C indicates a longer period than our observing baseline and correspondingly larger uncertainties in the orbital parameters. It is clear from our two imaging observations that B and C are also a parallax and proper motion pair since their relative separation changes by much less than the proper motion during the interval between the observations. The orbit is nearly edge on (Fig. 4). GJ 231.1BC may be relatively inactive with no reported H$\alpha$ or X-ray emission. It is also a member of the local sample based upon its distance, 18.1 pc, and space velocities (Reid, Hawley, & Gizis 1995). It has ~solar metallicity, [Fe/H] = $-0.02 \pm 0.04$ (Bonfils et al. 2005).

We again use the magnitude-color relationships to estimate that between the components delta $V = 2.85 \pm 0.21$ mag. Table 5 shows the component magnitudes. The $V$-$K$ magnitudes are $4.83 \pm 0.09$ for GJ 231.1B and $6.17 \pm 0.21$ for C, resulting in types of M3.5 for B and M5 for C. The masses for this system are not well-constrained by the astrometry but we use the MLRs again for estimates in the following section.

### 3.3 Mass Estimation

In the cases of unresolved astrometric systems in which the secondary contributes significantly to the total light, such as the subjects of this paper, we do not observe the secondary mass fraction, *f*, directly, but rather the quantity *f-β*. When we resolve the systems with imaging we measure the secondary light fraction, *β*, but since the components generally have different colors, the measurement of *β* in the IR does not unambiguously yield *β* in *V*, where required. However, as shown in §3.1 the color-magnitude relationships can be used to extrapolate from *H* to *V*. These two systems illustrate our increasing abilities to make mass estimates. For G 78-28 we use the astrometry and the imaging to make estimates for the total and component masses. Table 6 shows the estimates based upon the measured parallax, apparent magnitudes, the measured *JHK* ratios, the inferred *V* ratio, and two current MLRs (eqns. 2-5 Henry et al. 1993 – "HMLR," Delfosse et al. 2000 – "DMLR"). The uncertainties in HMLR consist of both the dispersion that they provide and our measurement errors, while those in DMLR are only the latter. There is good agreement within each MLR among the masses determined from the visible and near-IR colors. The two MLRs agree well for the mass of the primary, $M_A = 0.35 \pm 0.08$ $M_\odot$, but DMLR consistently predicts higher masses for the secondary, for which we adopt, $M_B = 0.18 \pm 0.04$ $M_\odot$. However, the estimates largely overlap.

The dynamical mass (§3.1) and the masses derived from the MLRs overlap but the mean of the dynamical mass is higher. The intersection of the two methods yields our adopted total mass estimate of $0.565 \pm 0.055$ $M_\odot$. Taking the derived values for *f* (§3.1) we get $M_B = 0.195 \pm 0.021$, consistent with the dynamical estimate and the two MLRs. Then, $M_A = 0.370 \pm 0.034$. We also estimate the spectral types from the mass-type relation in Kirkpatrick & McCarthy (1994). The types are M2 and M4, for A and B, respectively. Note that both the colors and the mass-type relation yield an earlier type for G 78-28A by 0.5-1 subtype than its prior value.

We perform a similar analysis for GJ 231.1BC and show the results in Table 7. Again, although the MLR estimates overlap, the masses derived from HMLR are consistently 0.01-0.02 $M_\odot$ smaller than those derived from DMLR. For this system, $M_B = 0.25 \pm 0.06$ $M_\odot$ and $M_C = 0.12 \pm 0.02$ $M_\odot$. The spectral types are M3 and M5 for B and C, respectively, from the mass-type relation (Kirkpatrick & McCarthy 1994). As in the case of G 78-28, the mass-type relation yields a sub-type that is ~0.5 earlier than the color-type for the primary, but agrees well for the secondary.

At the current level of knowledge and because of the intrinsic scatter of luminosities and masses due to stellar properties such as age and metallicity, the fact that systems such as G 78-28AB and GJ 231.1BC are binary has little impact on the estimated masses (or spectral types) of the primaries alone. For example, the mass estimate for G 78-28A decreases as calculated in HMLR by only ~0.03 $M_\odot$, within the error, because of the small decrease in luminosity of this binary component. Similarly, GJ 231.1B decreases by only ~0.01 $M_\odot$. However, the total masses of the systems change more dramatically, from 0.42 $M_\odot$ to 0.52 $M_\odot$ for G 78-28AB, and from 0.29 $M_\odot$ to 0.37 $M_\odot$ for GJ 231.1BC, even if we use the larger mass estimates for the single stars derived from the near-IR MLRs. If all the light came from a single star in both systems, the near-IR MLR mass estimates would be consistently higher than those due to the visible for both HMLR and DMLR. This is because the primary accounts for a smaller fraction of the light in the IR than in the visible. Therefore the IR mass estimate drops further than the visible mass estimate for the primary. Conversely, if the star is incorrectly believed to be single, then there is more erroneously assigned light in the IR than in the visible leading to a higher mass estimate in the IR. This apparent discrepancy in the masses derived from MLRs is largely removed with the discoveries of binarity (Tables 6 and 7).

Our derived masses for the four components agree well with theoretical model estimates. Fig. 8 shows our stellar $M_K$ and mass values on a plot with models of Baraffe et al. (1998). The data are consistent with models with solar or lower metallicity ([M/H] = 0 or -0.5) and age $t \geq 1.6 \times 10^8$ yr. The models do not provide an age upper limit since they are not distinguishable for $t > 10^9$ yr and these parameters.

## 4.0 SUMMARY

We have discovered two new low mass binary systems with measurements of their astrometric motions. Our follow-up imaging observations resolved the systems in the near IR. With the combined data we have improved mass estimates for four additional low mass stars. Paths for further improvements are to lower the dispersion in the MLRs by using discoveries such as these to form larger homogeneous samples as their inputs, measure delta *V* independently, and sharpen our knowledge of the orbit by extending the baseline of astrometric observations or otherwise increasing their accuracy.

## Table 1. Previously Known Properties

|  | G 78-28 | GJ 231.1B |
|---|---|---|
| **RA (2000)**[a] | 03 17 12.24 | 06 17 10.65 |
| **Dec (2000)**[a] | +45 22 22.0 | +05 06 00.4 |
| **V**[b] | 12.39 | 13.27 |
| **J**[c] | 8.422 ± 0.023 | 9.088 ± 0.015 |
| **H**[c] | 7.865 ± 0.019 | 8.559 ± 0.039 |
| **K**[c] | 7.593 ± 0.013 | 8.267 ± 0.011 |
| **Type** | M 3 | M 3.5 |
| **Parallax**[d] (*mas*) | 66 ± 13 | 50 ± 10 |
| **Proper Motion**[e] (*mas* y$^{-1}$) | 264 ± 7 | 287 ± 22 |
| **Position Angle**[e] (deg) | 253 ± 3 | 304 ± 4 |

[a]Salim & Gould 2003, Reid, Lepine et al. 2005 [b]Weis 1988, Weistrop 1981, [c]2MASS, [d]Gliese & Jahrweiss 1991, van Altena, Lee & Hoffleit 1995, Reid, Hawley, & Gizis 1995, [e] Luyten 1979, CMC 1999, Salim & Gould 2003, Lepine et al. 2005

## Table 2. Derived Stellar and System Parameters

|  | G 78-28AB | GJ 231.1BC |
|---|---|---|
| **Relative Parallax** (*mas*) | 52.4 ± 0.1 | 53.2 ± 0.2 |
| **Absolute Parallax** (*mas*) | 54.4 ± 1.0 | 55.2 ± 1.0 |
| **Proper Motion** (*mas* y$^{-1}$) | 269.3 ± 0.3 | 270.2 ± 3.6 |
| **Position Angle** (deg) | 253.3 ± 0.2 | 309.8 ± 3.6 |
| **Period** (y) | 4.18 ± 0.03 | > 25.7 |
| **Total Mass** (M$_\odot$) | 0.53 ± 0.09 | 0.37 ± 0.07 |
| **Semi-Major Axis** (AU) | 2.19 ± 0.10 | > 6.4 |
| **Eccentricity**, *e* | 0.281 ± 0.030 | -- |
| **Inclination** (deg) | 78 ± 1 | 90 ± 3 |
| **Lon. Asc. Node**[a] (deg) | 4.5 ± 0.5 | -- |
| **Arg. of Periapse** (deg) | 254.5 ± 1.5 | -- |
| **Epoch** | 2002.015± 0.035 | -- |
| **Primary Mass, M$_{pri}$** (M$_\odot$) | 0.370 ± 0.034 | 0.25 ± 0.06 |
| **Secondary Mass, M$_{sec}$** (M$_\odot$) | 0.195 ± 0.021 | 0.12 ± 0.02 |

[a]or + 180º because of into or out of plane ambiguity

## Table 3. LGSAO Measurements[a]

| Binary | Band | dMag | Sep. (*mas*) | PA (deg) |
|---|---|---|---|---|
| **G 78-28AB** | *J* | 1.24 ± 0.07 | | |
| **G 78-28AB** | *H* | 1.24 ± 0.07 | 70.4 ± 2.5 | 172 ± 2 |
| **G 78-28AB** | *Kp* | 1.14 ± 0.06 | | |
| **GJ 231.1BC** | *J* | 1.65 ± 0.05 | | |
| **GJ 231.1BC** | *Kp* | 1.52 ± 0.05 | 366 ± 3 | 158 ± 1 |

[a]JD=2453376.0

### Table 4. Palomar AO Measurements

| Binary | Julian Date | Band | dMag | Sep. (mas) | PA (deg) |
|---|---|---|---|---|---|
| G 78-28AB | | | | | |
| G 78-28AB | 2453004.8 | $H$ | [a] | 58.9 ± 1.3 | 32.3 ± 1.1 |
| G 78-28AB | 2453632.9 | $H$ | 1.285± 0.023 | 96.1 ± 1.1 | 184.9 ± 0.7 |
| G 78-28AB | 2453779.7 | $H$ | [a] | 41.8 ± 0.6 | 206.6 ± 0.9 |
| GJ 231.1 BC | 2453779.7 | $H$ | 1.64 ± 0.06 | -- | -- |
| GJ 231.1BC | 2453779.7 | $K_s$ | 1.53 ± 0.04 | 431 ± 4 | 157.9 ± 0.7 |

[a]fixed at the G 78-28 JD 2453632.9 value

### Table 5. Component Magnitudes

| | G 78-28A | G 78-28B | GJ 231.1B | GJ 231.1C |
|---|---|---|---|---|
| $M_V$ | 11.13 ± 0.05 | 12.98 ± 0.27 | 12.05 ± 0.07 | 14.90 ± 0.21 |
| $M_J^a$ | 7.40 ± 0.03 | 8.64 ± 0.07 | 8.01 ± 0.02 | 9.66 ± 0.05 |
| $M_H^a$ | 6.83 ± 0.03 | 8.12 ± 0.07 | 7.48 ± 0.04 | 9.12 ± 0.06 |
| $M_K^a$ | 6.59 ± 0.02 | 7.73 ± 0.06 | 7.22 ± 0.01 | 8.73 ± 0.05 |

[a]based on 2MASS and Tables 3 and 4

### Table 6. G 78-28AB Masses Derived from MLRs

| | Mass[a] ($M_\odot$) | | | Mass[b] ($M_\odot$) | | |
|---|---|---|---|---|---|---|
| Band | G 78-28A | G 78-28B | Total | G 78-28A | G 78-28B | Total |
| $V$ | 0.34±0.08 | 0.17±0.07 | 0.51±0.11 | 0.35±0.01 | 0.20±0.02 | 0.55±0.03 |
| $J$ | 0.38±0.10 | 0.18±0.03 | 0.56±0.11 | 0.35±0.02 | 0.21±0.01 | 0.55±0.03 |
| $H$ | 0.34±0.08 | 0.17±0.04 | 0.51±0.08 | 0.35±0.02 | 0.19±0.01 | 0.54±0.03 |
| $K$ | 0.34±0.07 | 0.18±0.04 | 0.52±0.08 | 0.35±0.02 | 0.20±0.01 | 0.55±0.02 |
| $V-K$ | -- | -- | -- | 0.35 ± 0.01 | 0.20 ± 0.03 | 0.50 ± 0.04 |
| Combined | 0.35±0.08 | 0.18±0.04 | 0.53±0.09 | 0.35±0.02 | 0.20±0.02 | 0.54±0.03 |

[a]based on Henry et al. (1993), [b]based on Delfosse et al. (2000)

### Table 7. GJ 231.1BC Masses Derived from MLRs

| | Mass[a] ($M_\odot$) | | | Mass[b] ($M_\odot$) | | |
|---|---|---|---|---|---|---|
| Band | GJ 231.1B | GJ 231.1C | Total | GJ 231.1B | GJ 231.1C | Total |
| $V$ | 0.25±0.05 | 0.11±0.02 | 0.37±0.06 | 0.28±0.01 | 0.12±0.02 | 0.41±0.03 |
| $J$ | 0.26±0.07 | 0.12±0.02 | 0.38±0.09 | 0.27±0.01 | 0.13±0.05 | 0.39±0.02 |
| $H$ | 0.24±0.05 | 0.12±0.02 | 0.36±0.07 | 0.27±0.01 | 0.13±0.01 | 0.40±0.02 |
| $K$ | 0.24±0.05 | 0.12±0.02 | 0.36±0.07 | 0.26±0.01 | 0.13±0.01 | 0.38±0.01 |
| $V-K$ | -- | -- | -- | 0.27 ± 0.01 | 0.13 ± 0.01 | 0.40 ± 0.02 |
| Combined | 0.25±0.06 | 0.12±0.02 | 0.37±0.07 | 0.27±0.01 | 0.13±0.02 | 0.40±0.02 |

[a]based on Henry et al. (1993), [b]based on Delfosse et al. (2000)

**Appendix: Plate Scale**

The LGSAO system provides a unique opportunity to determine plate scales for NIRC2. The core of the globular cluster M5 was observed on the night of April 30, 2005. A total of 400 *Kp* band, co-added images of 0.2 second each were taken in the narrow and wide cameras. The images were dithered, and Antonin Bouchez' IDL routine nirc2warp.pro was used to remove known distortion from the NIRC2 detector[2] before the images were combined into mosaics. The resulting mosaics are roughly 15" and 61" across for the narrow and wide cameras, respectively. The positions of 17 stars were compared between mosaics from both camera modes as well as from the Wide Field Planetary Camera 2 (WFPC2) aboard the Hubble Space Telescope. The stars span the area of the mosaics. The public, 96-second WFPC2 image was taken by F. R. Ferraro (proposal 6607) in the F555W filter on July 26, 1997 and retrieved using the Multimission Archive at Space Telescope[3]. The four bright stars in the center of the fields were not used, as they saturated the NIRC2 detector.

In the WFPC2 image, the core of M5 lies on the high resolution, Planetary Camera chip. Holtzman, *et al*. (1995), derive the Planetary Camera pixel scale as 45.54 mas/pix from observation of "several dozen" stars in the core of the globular cluster ω Cen, and they calculate the pixel scale as 45.55 mas/pix from observation of 11 stars in the globular cluster M67. No errors are given for their pixel scale; we therefore assume a Planetary Camera pixel scale of 45.545 ± 0.005 mas/pix.

Because of the higher resolution of the NIRC2 mosaics, we assume that stellar centroiding is more accurate than in the Planetary Camera image. Therefore, we treat the NIRC2 mosaics as our reference images (described below). We fit second-order polynomials, in both x- and y-axes, to the stellar positions between each reference image and the Planetary Camera image. We allow the pixel scales in the x- and y-directions to differ from each other. Our transformations are given by

(1a) $$x' = ax^2 + bx + cy^2 + dy + exy + f,$$
(1b) $$y' = gy^2 + hx + iy^2 + jy + kxy + l,$$

where the primed coordinates are stellar positions in the Planetary Camera image, and the unprimed coordinates are positions in the NIRC2 reference images. Having accurate centroids for stars in the reference images is necessary for accurate transformations. The relative linear scalings, in the x- and y-directions, between each NIRC2 mosaic and the Planetary Camera image are thus

(2a) $$m_x = \sqrt{\left(\frac{\partial x'}{\partial x}\right)^2 + \left(\frac{\partial x'}{\partial y}\right)^2},$$

(2b) $$m_y = \sqrt{\left(\frac{\partial y'}{\partial x}\right)^2 + \left(\frac{\partial y'}{\partial y}\right)^2}.$$

---

[2] http://www2.keck.hawaii.edu/optics/lgsao/software/nirc2warp.pro
[3] http://archive.stsci.edu/

Because of the second-order terms in eqns. 1, relative scaling between the reference and Planetary Camera images is not independent of stellar position. This is due to errors in the mosaicking process as well as residual distortion present in the detector. These errors limit the accuracy to which pixel scale can be measured; observing a denser star field, for example, will not increase the accuracy of this measurement. The x- and y-pixel scales for the NIRC2 detector are therefore given by

$$s_x = \overline{m}_x s'_x, \tag{3a}$$
$$s_y = \overline{m}_y s'_y. \tag{3b}$$

Note that $\overline{m}$ represents the mean relative scaling calculated from all 17 stellar positions in the reference image. The primed quantities are on the opposite sides of the transformation quantities between eqns. 1 and 3; this is because the units of eqns. 1 are in pixels, while the units on eqns. 3 are per pixel. Uncertainties in pixel scale are calculated according to

$$\sigma_{s_x} = \sqrt{\left(\sigma_{m_x} s'_x\right)^2 + \left(\overline{m}_x \sigma_{s'_x}\right)^2}, \tag{4a}$$
$$\sigma_{s_y} = \sqrt{\left(\sigma_{m_y} s'_y\right)^2 + \left(\overline{m}_y \sigma_{s'_y}\right)^2}, \tag{4b}$$

where $\sigma_m$ represents the standard deviation of the relative scaling.

The PAs of each pair of stars were compared between mosaics, and this allows the net rotation between each pair of mosaics to be determined. Since the header keywords for each mosaic were used to determine North on the detectors, the net rotation represents the accuracy of those keywords. Table A1 lists the x- and y-pixel scales for both the NIRC2 narrow and wide cameras, and Table A2 compares these with values in the literature. The net pixel scales in this work comprise the mean of the x and y pixel scales, and the error is half the quadrature addition of the x and y pixel scale errors. Finally, Table A3 shows the net rotation between each pair of mosaics.

Note that our x and y pixel scales fall within the error bounds from the literature. Our narrow camera x and y pixel scales agree to one standard deviation.

### Table A1: NIRC2 Pixel Scales

| NIRC2 mode | x pixel scale (*mas*) | y pixel scale (*mas*) |
|---|---|---|
| Narrow | 9.982 ± 0.024 | 9.958 ± 0.012 |
| Wide | 39.905 ± 0.076 | 39.862 ± 0.019 |

### Table A2: Comparison to Literature

| Group | Narrow (*mas*/pix) | Wide (*mas*/pix) |
|---|---|---|
| Ghez, *et al.* | 9.93 ± 0.05 | — |
| König, *et al.* | 9.942 ± 0.500 | — |
| Roe, *et al.* | 9.95 ± 0.02 | — |
| This work | 9.970 ± 0.012 | 39.884 ± 0.039 |

### Table A3: Accuracy of Rotation Keywords

| Pair of mosaics | Net rotation (°) |
|---|---|
| Narrow to Wide | 0.026 ± 0.087 |
| PC to Narrow | 0.133 ± 0.074 |
| PC to Wide | 0.158 ± 0.089 |

**ACKNOWLEDGMENTS**
The research described in this paper was performed in part by the Jet Propulsion Laboratory, California Institute of Technology, under contract with the National Aeronautics and Space Administration. We performed observations at Caltech's Palomar Observatory and Keck Observatory and acknowledge the assistance of the staff. We thank I. Baraffe and G. Chabrier for their assistance. This research has made use of the NASA/IPAC Infrared Science Archive, which is operated by the Jet Propulsion Laboratory, California Institute of Technology, under contract with the National Aeronautics and Space Administration. This research has made use of the SIMBAD database, operated at CDS, Strasbourg, France, and of NASA's Astrophysics Data System Abstract Service. This publication makes use of data products from the Two Micron All Sky Survey, which is a joint project of the University of Massachusetts and the Infrared Processing and Analysis Center/California Institute of Technology, funded by the National Aeronautics and Space Administration and the National Science Foundation. M.I. acknowledges Michelson Fellowship support from the Michelson Science Center and the NASA Navigator Program.

**FIGURE CAPTIONS**

1. The astrometric motion of G 78-28 measured with STEPS. The reference JD is 2450803.78. The model has a orbital period, P = 4.19 y, eccentricity, $e = 0.26$, system mass = 0.57 $M_\odot$, and semi-major axis, $a = 2.15$ AU.

2. The G 78-28 astrometric data (points) superimposed on the orbital model (curve) with the parameters listed in the Fig. 1 caption. The data are from STEPS (black diamonds), LGSAO (red square), and Palomar AO (blue circles). The Palomar AO measurement errors are the size of the symbols.

3. GJ 231.1BC allowed orbital periods vs. eccentricity. The total system mass is restricted to the value and uncertainty determined with the MLRs.

4. A possible orbit for GJ 2311.1BC. The data are from STEPS (black diamonds), LGSAO (red square), and Palomar AO (blue circles). The model is for a orbit with P = 102 y, eccentricity, $e = 0.43$, system mass = 0.37 $M_\odot$, and semi-major axis, $a = 15.6$ AU.

5. G 78-28A and B in the *H* band. North is up and East is to the left.

6. GJ 231.1B and C in the *Kp* band. North is up and East is to the left.

7. The correlation between the derived parameters, brightness ratio and separation, for the Palomar AO observations of G 78-28 on JD 2453779.7 (a, left) and 2453779.7 (b, right). The contours show confidence regions of 90, 99, and 99.9%.

8. A plot of our data and the models of Baraffe et al. (1998). The models have solar metallicity with ages: $t = 10^8$ yr (dotted line), $t = 1.6 \times 10^8$ yr (dashed line), and $t = 10^9$ yr (solid line). Our data points are labeled.

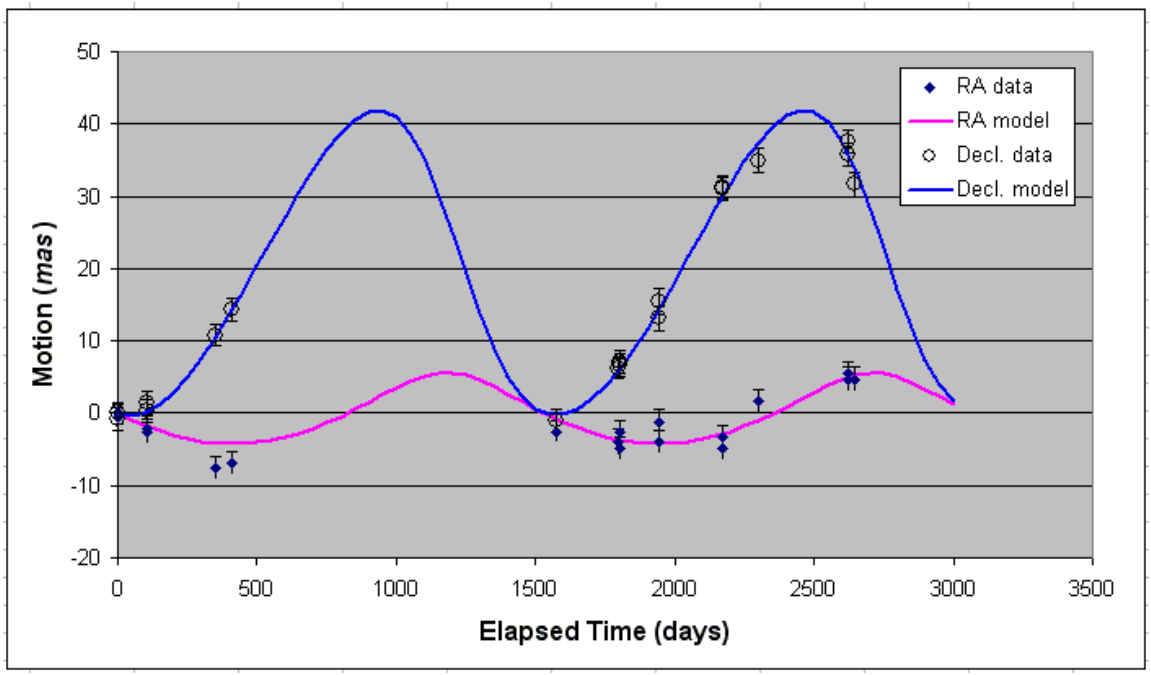

**Figure 1.**

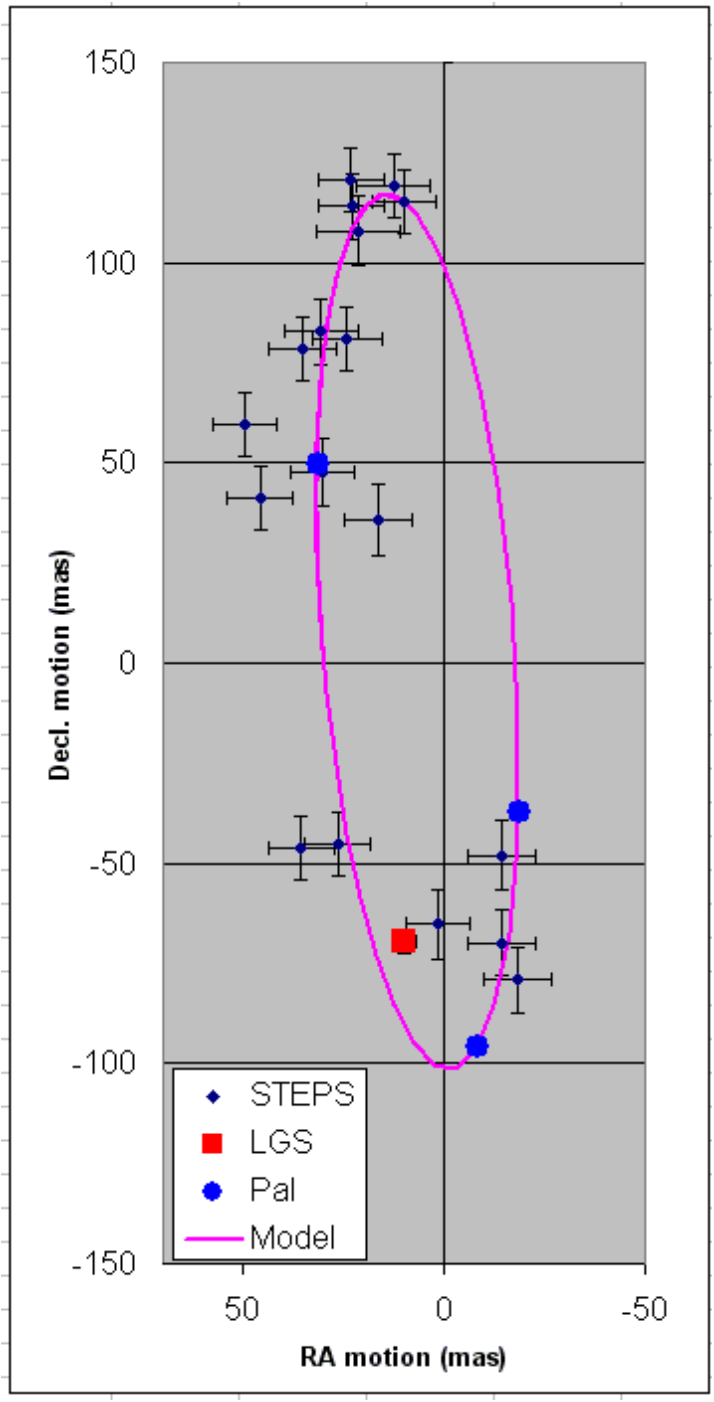

**Figure 2.**

**Figure 3**

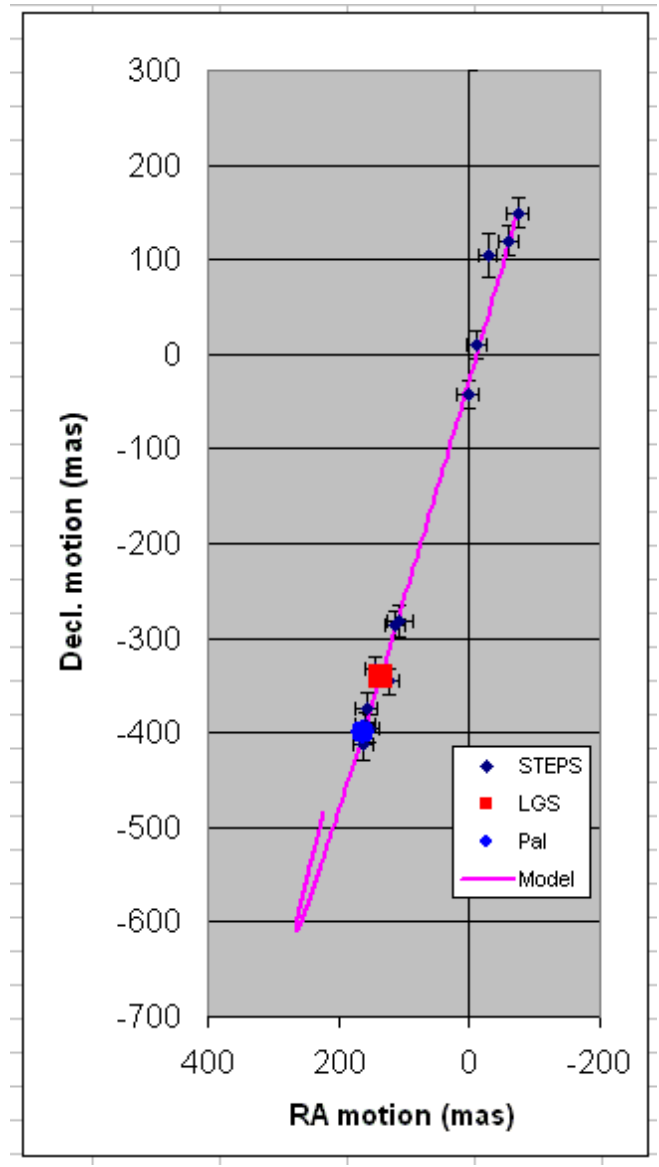

**Figure 4.**

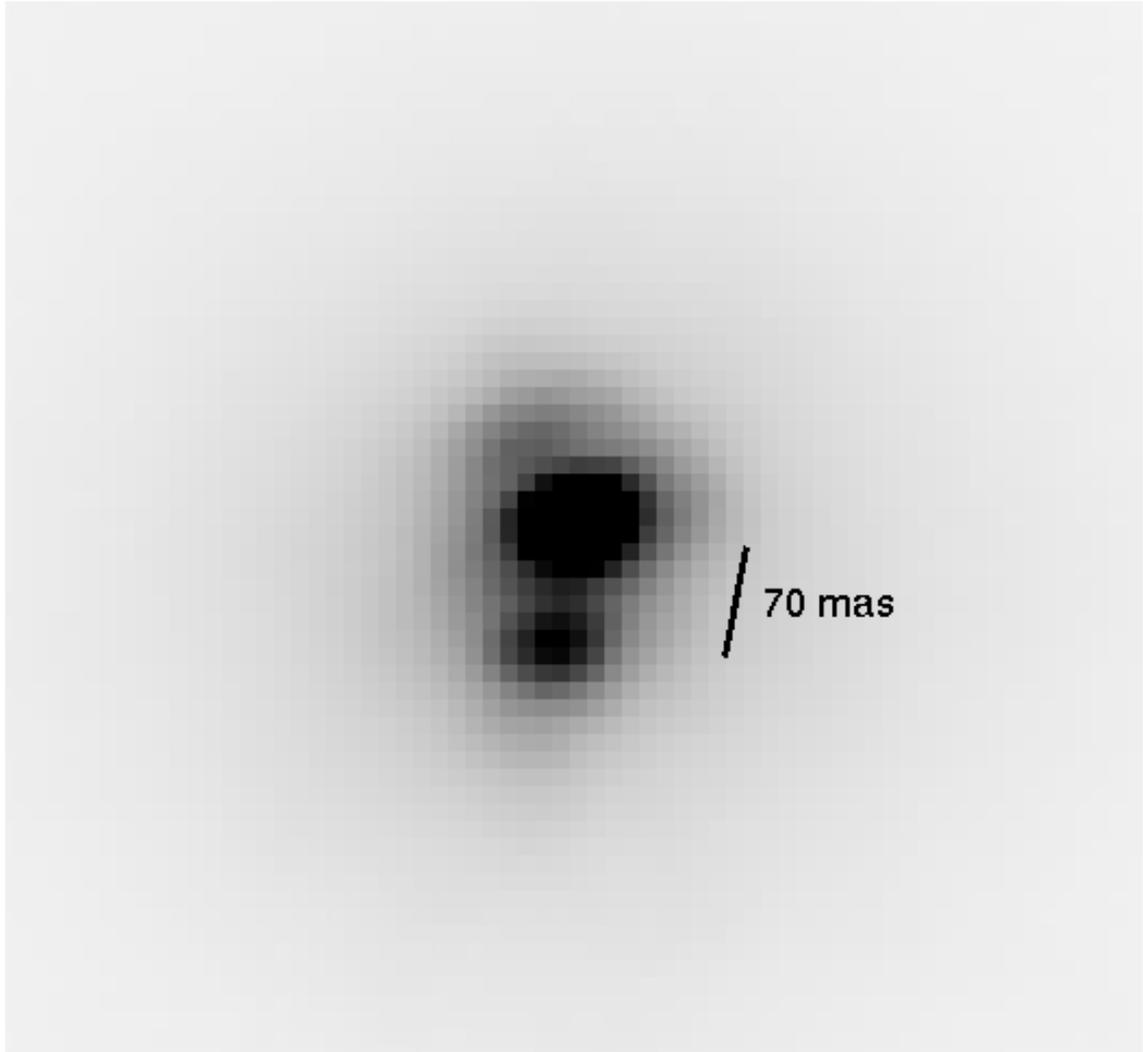

**Figure 5.**

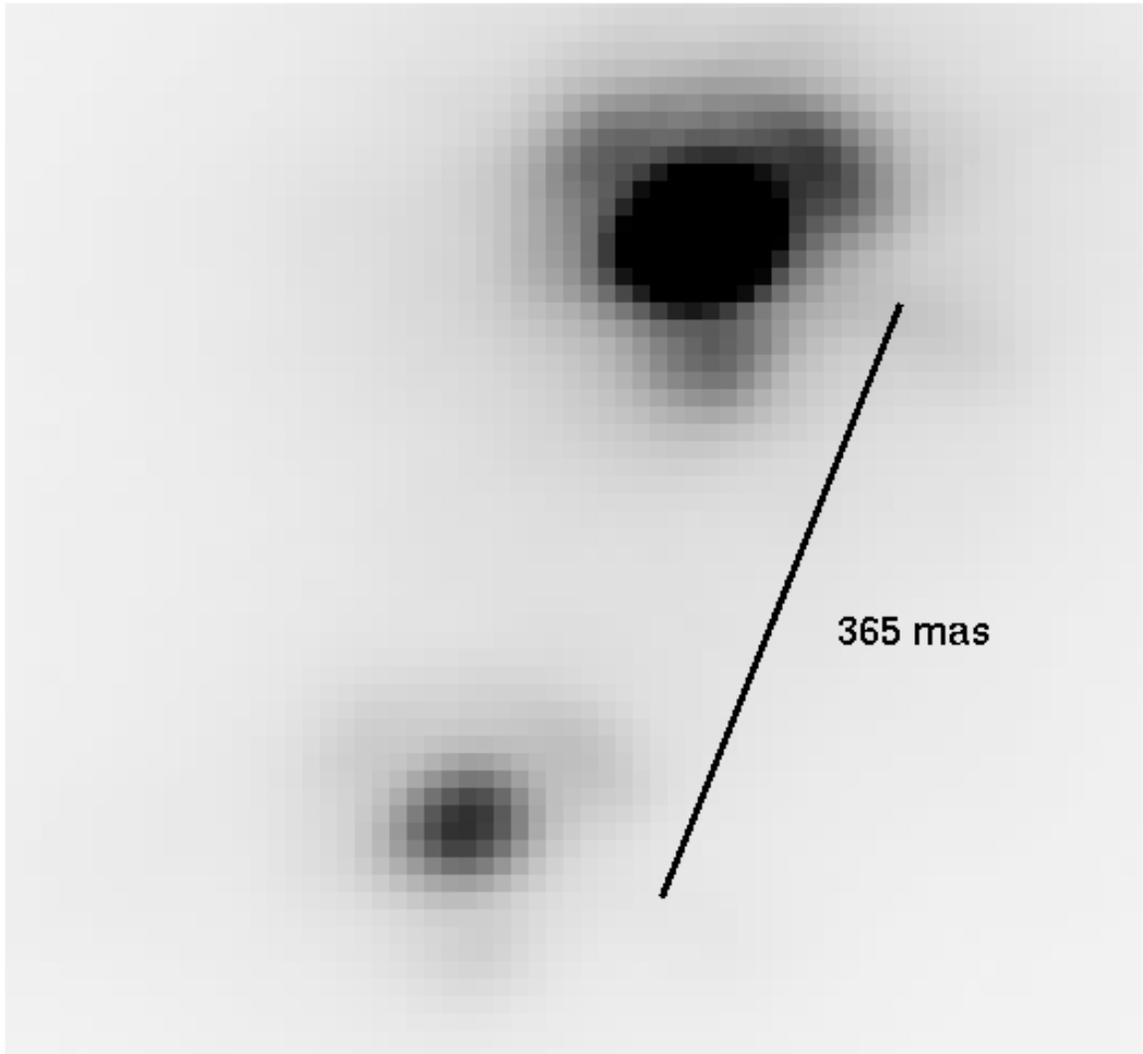

**Figure 6.**

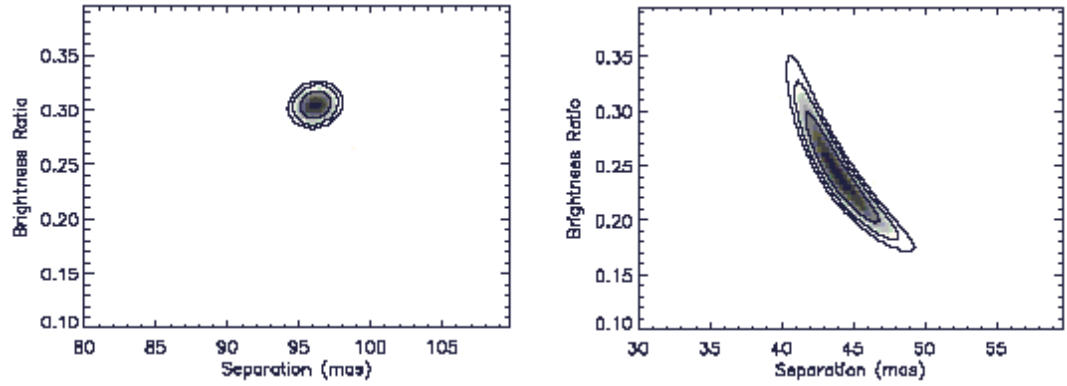

**Figure 7**

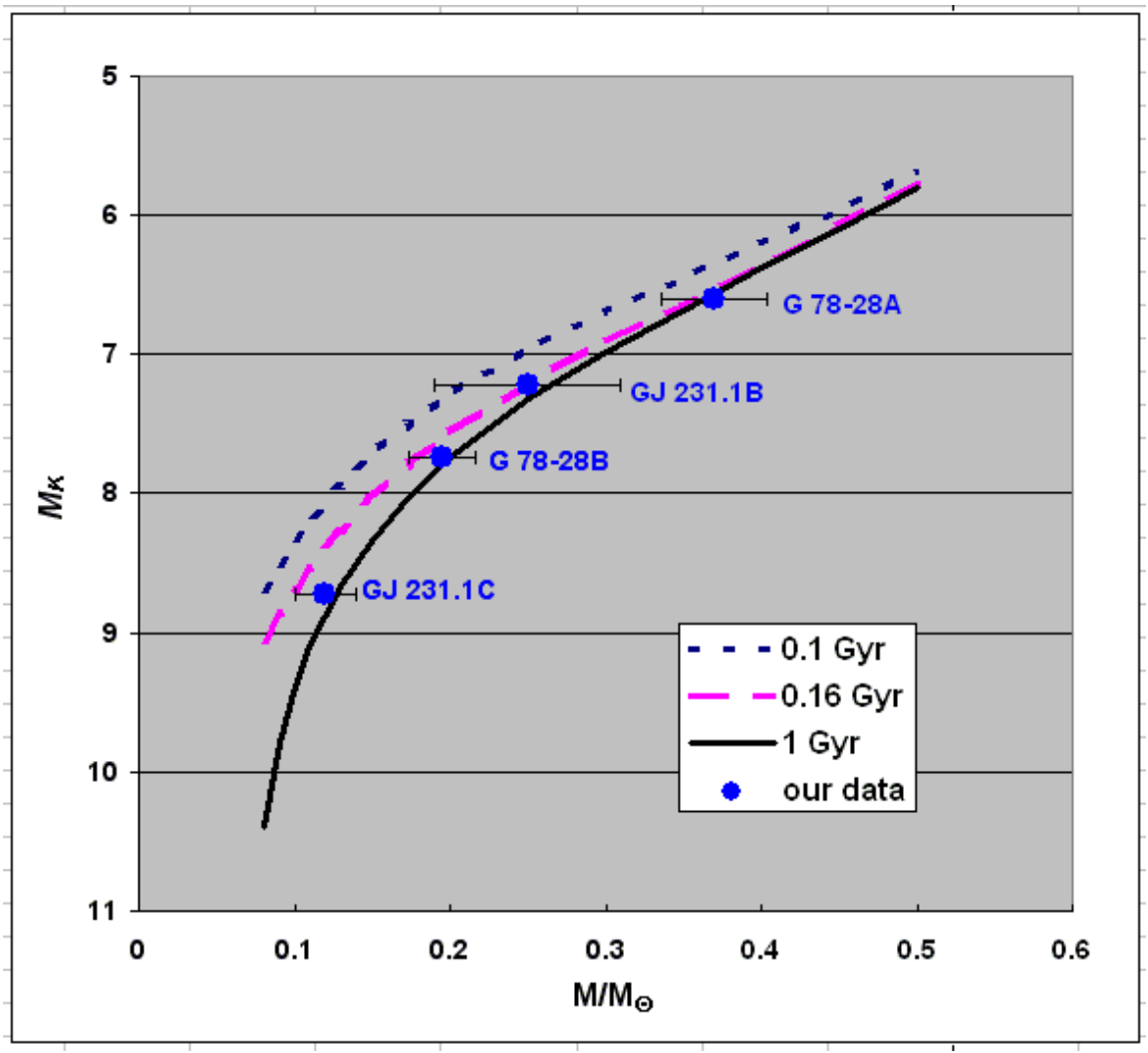

**Figure 8**

25